\documentclass{article}
\usepackage{spconf,amsmath,graphicx}
\graphicspath{{./figs/}}
\usepackage{amsthm}

\theoremstyle{definition}

\theoremstyle{remark}

\usepackage{array}
\usepackage{cite}

\usepackage{pgfplots}

\usepackage{epstopdf}
\usepackage{marginnote}
\usepackage{mathtools}
\usepackage{pseudocode}
\usepackage{rotating}

\usepackage{lipsum}                     
\usepackage{xargs}                      

\usepackage{setspace}
\usepackage{xr}
\usepackage{amssymb}
\usepackage{epstopdf}
\usepackage{color}
\usepackage{amsthm}
\usepackage{caption}

\newcommand{\mbf}[1]{\mathbf{{#1}}  }
\usepackage{comment}
\graphicspath{{./figs/}}

\title{Dynamic MRI using deep manifold self-learning}

\name{Abdul Haseeb Ahmed, Hemant Aggarwal, Prashant Nagpal, Mathews Jacob}
\address{Department of Electrical and Computer Engineering, University of IOWA}
\begin{document}
%
\maketitle
\begin{abstract}
We propose a deep self-learning algorithm to learn the manifold structure of free-breathing and ungated cardiac data and to recover the cardiac CINE MRI from highly undersampled measurements. Our method learns the manifold structure in the dynamic data from navigators using autoencoder network. The trained autoencoder is then used as a prior in the image reconstruction framework. We have tested the proposed method on free-breathing and ungated cardiac CINE data, which is acquired using a navigated golden-angle gradient-echo radial sequence. Results show the ability of our method to better capture the manifold structure, thus providing us reduced spatial and temporal blurring as compared to the SToRM reconstruction.
\end{abstract}
\begin{keywords}
Cardiac MRI, denoising auto-enocoder, deep learning, image reconstruction
\end{keywords}
\section{Introduction}
\label{sec:intro}
Dynamic MRI (DMRI) plays a central role in several clinical applications such as cardiac CINE MRI, which is widely used in clinics for the anatomical and functional assessment. The clinical practice is to acquire the CINE data using breath holding to achieve good spatial and temporal resolution. However, it is difficult for many subjects, including children, patients with myocardial infarction, and chronic obstructive pulmonary disease patients to hold their breath. In addition, multiple breath-holds prolong the scan time, adversely impacting patient comfort and compliance. 

Several acceleration methods, including parallel MRI and compressed sensing, are currently exploited to reduce the breath-hold duration in cardiac CINE applications. Subspace models\cite{ktfocuss,ktslr} that learn from the data itself can represent the pixel intensity profiles compactly. This representation facilitates the recovery of DMRI data from highly undersampled measurements. Recently, non-linear manifold models that rely on kernel low-rank relation have been shown to outperform classical subspace based models in the context of free-breathing and ungated cardiac MRI \cite{sunritatci,leslieklr}. These non-linear models are observed to be more efficient in representing the dynamics of both cardiac and respiratory motions. Many of these schemes rely on k-space navigator measurements, which are acquired from specific k-space locations at each temporal frame to estimate the subspace. The estimated subspace model is then used to recover the dynamic dataset from highly undersampled measurements. While these methods offer great potential, even higher acceleration is needed to extend these applications to 3D+time setting. 

Deep learning algorithms are now emerging as powerful alternatives to recover MRI data from undersampled measurements. Most of the current methods rely on convolutional neural networks (CNNs), which learn the manifold structure of patches from fully sampled exemplary images. These pre-learned models are then used to recover MRI data from undersampled measurements. Both direct inversion methods that rely on a large CNN to recover the data directly from the measurements as well as model-based strategies that rely on regularization penalties that involve smaller CNN blocks, have been introduced with promising results. By contrast, not much work is done on learning the redundancy in the temporal profiles in DMRI. 

\begin{figure*}[t!]
	\centering
	\includegraphics[ scale=0.8]{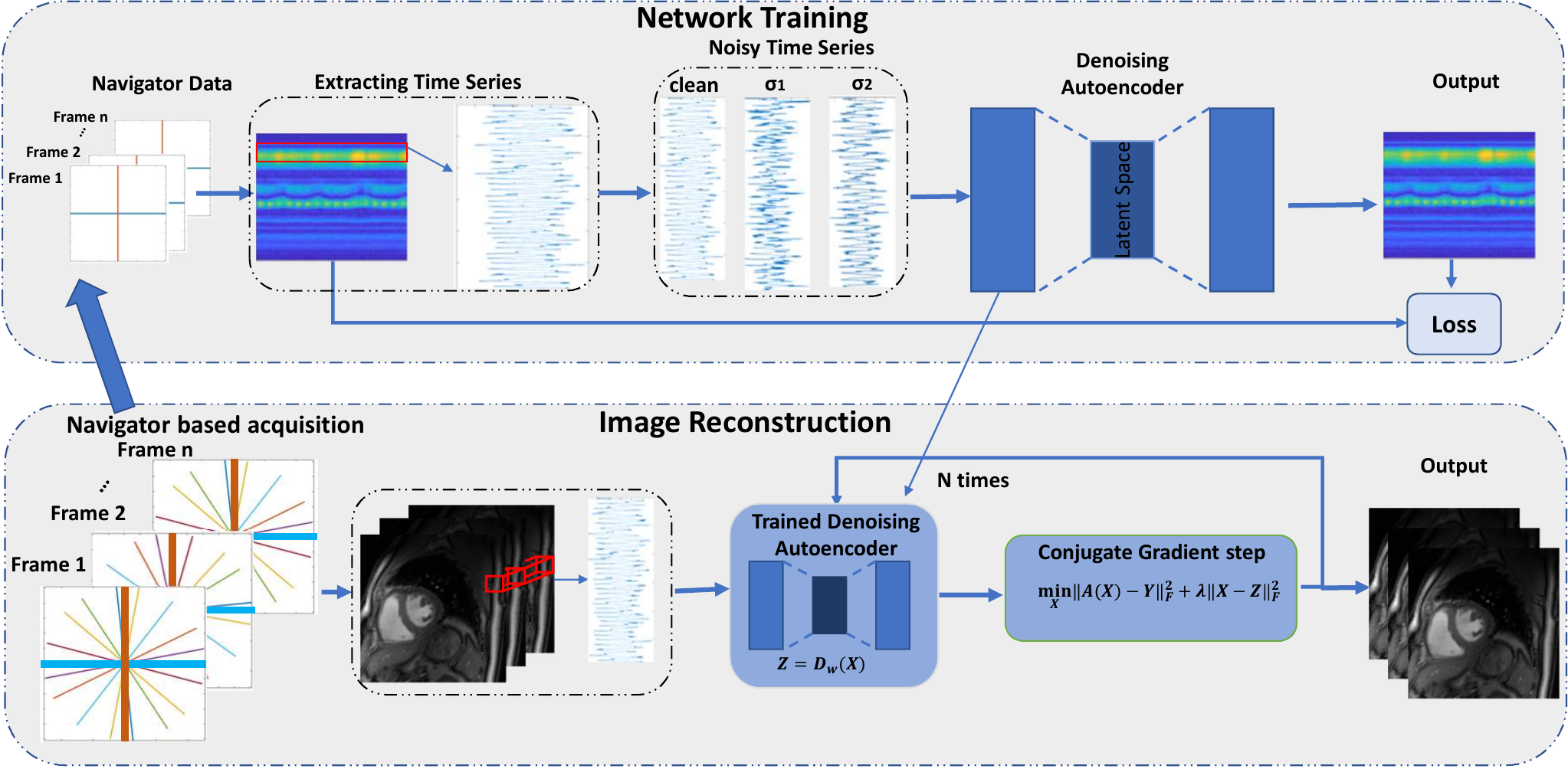}\\
	\caption {\small {Proposed method: Top row shows the training of the DAE using the navigator data $\mathbf Z$. Specifically, the navigator signals are corrupted by Gaussian noise and fed into the network, whose parameters are learned to denoise the navigators. Once the denoiser is trained, it is used within the reconstruction network as shown in the bottom row, where the residual error of the DAE is used as a prior. }}
	\label{fig1}
\end{figure*}  

The main focus of this work is to introduce a deep learning algorithm to self-learn the temporal dynamics of the dataset from its k-space navigators. The self-learned network is then used to recover the DMRI dataset from highly undersampled measurements. This approach is thus conceptually similar to the partial separable function (PSF) framework that exploits the subspace structure of DMRI data. We rely on a denoising auto-encoder (DAE) in this work. We are motivated by theoretical results, which shows that DAE captures the smoothed probability density of the data, while the residual representation error is a measure of the derivative of the log density \cite{vincent08}. We use the residual error in representing the voxel profiles as the regularization penalty in a model-based reconstruction framework. This approach can be viewed as the generalization of classical low-rank subspace prior, where norm of the error is chosen as the penalty that projects the voxel profiles to the data manifold. We compare the proposed scheme against the SToRM approach, which relies on kernel low-rank regularization. 

We note that the self-learning strategy has conceptual similarities to the RAKI framework \cite{raki}, where self-learned deep networks are used to interpolate parallel MRI data in k-space. The focus of this work is on DMRI, which is fundamentally different from the RAKI setting. Unlike RAKI, we use a model-based framework, where the SENSE forward model is used along with the self-learned prior.

\section{METHOD}
\label{sec:format}
\subsection{Navigator based MRI acquistion}
We model the multichannel acquisition of the dynamic data $\mbf x(\mbf r,t):\mathbb{Z}^3 \rightarrow \mathbb{C}$ as:
\begin{equation}
\mbf y_i(\mathbf k,t) = \int_{\mathbf r} \mbf x(\mathbf r,t)~s_i(\mathbf r)e^{\,j(\mathbf k^T \mathbf r)}\,d\mathbf r+\mbf n(\mathbf r,t)
\end{equation}
Here, $\mathbf r = (x,y)$ and $\mathbf k=(k_x,k_y)$ represent the spatial and k-space locations, respectively. 
The above equation can be compactly rewritten in matrix form as:
\begin{equation}
\label{measures}
\mathbf Y = \mathcal A (\mathbf X) + \mathbf N,
\end{equation}
where $\mathcal A$ is the multi-channel undersampling forward model. Here $\mbf X \in \mathbb C^{m\times n}$ is the Casorati matrix of the data $\mbf x(\mathbf r,t)$. Here $m$ is the number of pixels in each time frame and $n$ is the total number of frames. In several acquisition schemes, it is a common practice to acquire each frame at the same k-space locations; this data is often called as k-space navigator signals. This process can be mathematically denoted as $\mathbf Z = \mathbf P \mathbf X$, where $\mathbf P$ is a linear measurement matrix.

\subsection{Low-rank/Subspace constrained dynamic MRI}
We now briefly review the low-rank/subspace approach, where the voxel time profiles are constrained to be in a subspace, to set the stage for the proposed scheme. These schemes express the data as $\mathbf X = \mathbf U \mathbf V^{H}$, where $\mathbf U\in \mathbb C^{m\times r}$ and  $\mathbf V\in \mathbb C^{n\times r}$. Specifically, each voxel profile is expressed as the weighted linear combination of $r$ basis functions specified by the rows of $\mathbf V$. The temporal basis functions $\mathbf V$ are often learned from k-space navigators $\mathbf Z$ by singular value decomposition. Once the temporal basis functions are determined, subspace constrained recovery is often posed as:
\begin{equation}
\label{lowrankclassical}
\mbf X =  \widehat{\mathbf U} \mathbf V^{H}, \mbox {  where } \widehat{\mathbf U} = \arg \min_{\mbf U \in \mathbb C^{m\times r}} \|\mathcal A (\mbf U\mathbf V^{H})-\mathbf Y\|^{2}_{2}
\end{equation}
\subsection{Reformulation of low-rank methods}
We now reformulate \eqref{lowrankclassical} using a penalized formulation as
\begin{equation}
\label{reformulate}
\mbf X^{*} =  \arg \min_{\mbf X \in \mathbb C^{m\times n}} \|\mathcal A (\mathbf X)-\mathbf Y\|^{2}_{2} + \lambda \|\mathbf X \underbrace{\left(\mathbf I_{n\times n} - \mathbf V \mathbf V^{H}\right)}_{\mathbf N}\|^{2}
\end{equation}
Here, $\mathbf N$ is a projection operator to the null-space; $\mathbf \|\mathbf X\mathbf N\|^{2}$ is the energy of the projection of $\mathbf X$ to the null-space. As $\lambda \rightarrow \infty$, the temporal profiles (columns of $\mathbf X$) will be constrained to be in the signal subspace and hence \eqref{reformulate} is equivalent to \eqref{lowrankclassical}. The subspace framework is well-suited to breath-held and low motion applications. When there is extensive motion, the voxel profiles may lie on a complex manifold and hence a linear subspace model may be inefficient in capturing the non-linear redundancies in the voxel profiles. 

Recently, kernel low-rank methods were introduced to exploit non-linear redundancies in the voxel profiles \cite{sunritatci}. These approaches rely on an optimization scheme similar to \eqref{reformulate} to recover the images. The main distinction is that the matrix $\mathbf N$ is derived using kernel low-rank optimization from navigator data. The derived $\mathbf N$ matrix is full rank and sparse, exploiting the non-linear structure in the voxel profiles. 

\subsection{Manifold constrained DMRI using DAE}
Motivated by the projection error based penalty in \eqref{reformulate}, we introduce a self-learning DMRI framework based on
denoising autoencoders (DAE).  DAEs were introduced as unsupervised schemes to learn the data manifold. Theoretical results show that the DAE representation error is a measure of the derivative of the smoothed log density \cite{vincent08} of the data; the derivative is zero if the point is on the manifold, while it is high when the point moves away from the \emph{data-manifold}. 

As shown in the first row of Fig. \ref{fig1}, we propose to self-learn the DAE parameters,  denoted by $\Theta$, from navigator data $\mathbf Z$ of each subject:
\begin{equation}\label{daetraining}
\Theta^* = \arg \min_{{\Theta}} \mathbb E_{I} \left(\mathbb E_{\mathbf S \sim  \mathcal N(\mathbf 0,\sigma_i^2)}\|\mathcal D_{\Theta}\left(\mathbf Z+\mathbf S\right) - \mathbf Z\|_F^2\right)
\end{equation}
Here, $\mathbb E$ denotes the expectation operator. Here $\mathbf S$ is a noise realization with a zero mean complex Gaussian density with variance $\sigma_{I}^{2}$; the $\sigma_{i}$ are chosen from a set of variances, indexed within the set $I$. Once the DAE parameters are self learned, we recover the time-series from its highly undersampled measurements \eqref{measures} by solving the optimization problem.
\begin{equation}
\mbf X^{*} = \arg \min_{\mbf X} \|\mathcal A (\mbf X)-\mathbf Y\|^{2}_{2}+ \lambda~ \|\mathbf X - \mathcal D_{\Theta}(\mathbf X)\|^{2},
\end{equation}
where $\mathbf X-\mathcal D_{\Theta}(\mathbf X)$ is the DAE error. We rely on an alternating minimization algorithm that alternates between the following steps as shown in the second row of Fig. \ref{fig1}:
\begin{eqnarray*}
	\mathbf X_{n+1} &=&  \arg \min_{\mbf X} \|\mathcal A (\mbf X)-\mathbf Y\|^{2}+ \lambda~ \|\mathbf X-\mathbf Q_n\|^2\\
	\mathbf Q_{n+1} &=& \mathcal D_{\Theta}(\mathbf X_{n+1})
\end{eqnarray*}

\subsection{Data acquisition and post-processing}
The data was acquired by a golden angle FLASH sequence on a Siemens 1.5T scanner (skyra) with 34 coil elements total (body and spine coil arrays) in the free-breathing and ungated mode from cardiac MRI patients with a scan time of 42 seconds per slice; the study was an add-on to the routine cardiac MRI exams.  Each frame was sampled by two k-space navigator spokes, oriented at 0 degrees and 90 degrees respectively. The protocol was approved by the Institutional Review Board (IRB) at the University of Iowa. The sequence parameters were: TR/TE 4.68/2.1 ms, FOV 300 mm, base resolution 256, slice  thickness 8 mm. A temporal resolution of 46.8 ms was obtained by sampling 10 lines of k-space per frame. The scan parameters were kept same across all patients. To reduce the computational complexity, we combined the data from $34$ channels to $7$ using principal component analysis. For the experiments in this work, we retained the initial $19$ seconds of the original acquisition. The reconstructed results were compared to SToRM reconstructions from $43$ seconds of data as ground truth. Both SToRM results relied on the Laplacian matrix estimation scheme in \cite{sunritatci}.

The radial k-space navigator signals from the dataset were inverse Fourier transformed and fed into the auto-encoder for training, as illustrated in Fig. \ref{fig1}. We used a DAE with four fully connected layers with RELU activation. Since we are using 400 frames, the input to the network has a dimension of 400, while the bottle neck layer has 50 features. We corrupted the input data with different noise realizations, each with different noise levels (10\%, 5\%, 3\%, 1\%, 0.7\%, 0.5\%, 0.3\%, and 0.1\%) and trained. More noise realizations with lower levels of noise were chosen to encourage the network to learn the identity mapping on the manifold. 
\begin{figure}[htb]
	\centering\vspace{-.5em}
	\includegraphics[ scale=0.6]{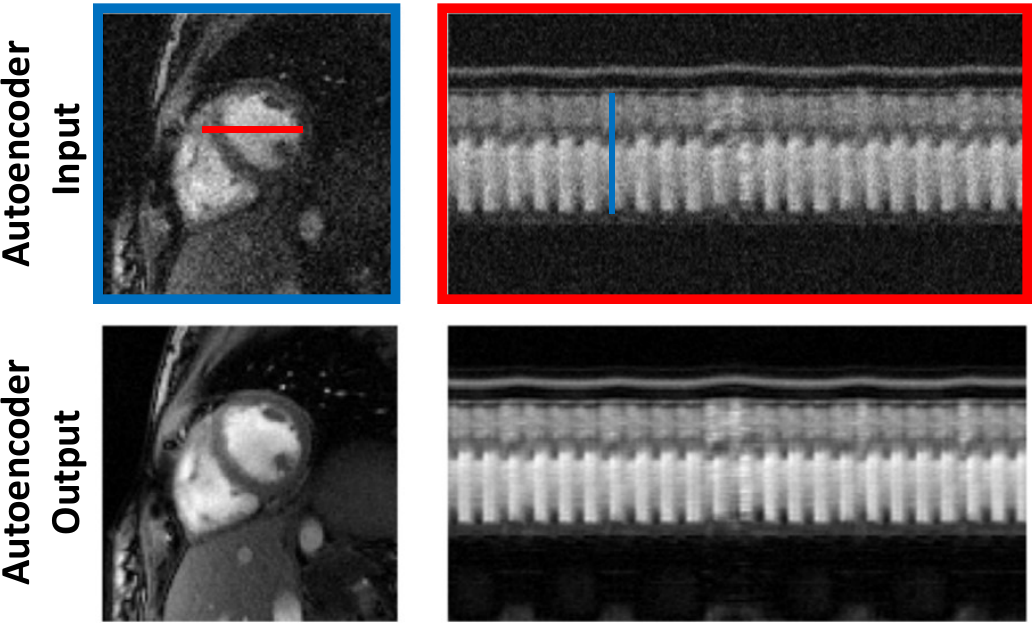}\\
	\caption {\small {Illustration of the denoising performance of the learned DAE. Top row: Input noisy image and its time profile from all the frames along red line. Bottom row: Output of the DAE, which shows the denoised images with preserved temporal features.}}\vspace{-3em}
	\label{fig3}
\end{figure}
 \begin{table}[]
	\begin{center}
	\scriptsize
	   \captionsetup{justification=centering}
		\caption{\\ \small Quantitative comparison of the methods on CINE data. \cite{sunritatci}}.
		\label{table1}
		\begin{tabular}{|c|c|c|c|c|c}
		\hline
		& Method & SER & SSIM & HFEN\\ \hline
		Data 1& SToRM& ${13.6 \pm 0.5}$    & ${0.76 \pm 0.02}$          & ${0.44 \pm 0.02}$ \\ \hline
		Data 1&Proposed&  $\bold{15.9 \pm 0.5}$    & $\bold{0.83 \pm 0.04}$          & $\bold{0.38 \pm 0.02}$ \\ \hline
		Data 2& SToRM& ${9.8 \pm 0.4}$    & ${0.69 \pm 0.03}$          & ${0.58 \pm 0.05}$   \\ \hline
		Data 2&Proposed&  $\bold{16.1 \pm 0.7}$    & $\bold{0.80 \pm 0.03}$          & $\bold{0.39 \pm 0.02}$   \\ \hline
		\end{tabular}\vspace{-2em}
	\end{center}		
\end{table}

\section{Experiments \& Results}
\label{sec:pagestyle}
We first test the denoising ability of our proposed scheme on the data recovered by SToRM method in Fig \ref{fig3}. The top row shows the noisy input image with its temporal profile. Bottom row shows the output of the autoencoder. We observe the network is capable of denoising the images, while preserving fine details and temporal fidelity.

Table \ref{table1} shows the quantitative comparisons with the 900 frame SToRM reconstructions as the ground truth. The proposed method gives better signal to error ratio (SER), structural similarity index (SSIM) and high frequency error (HFEN). 

\begin{figure}[t!]
	\centering
	\includegraphics[ scale=0.75]{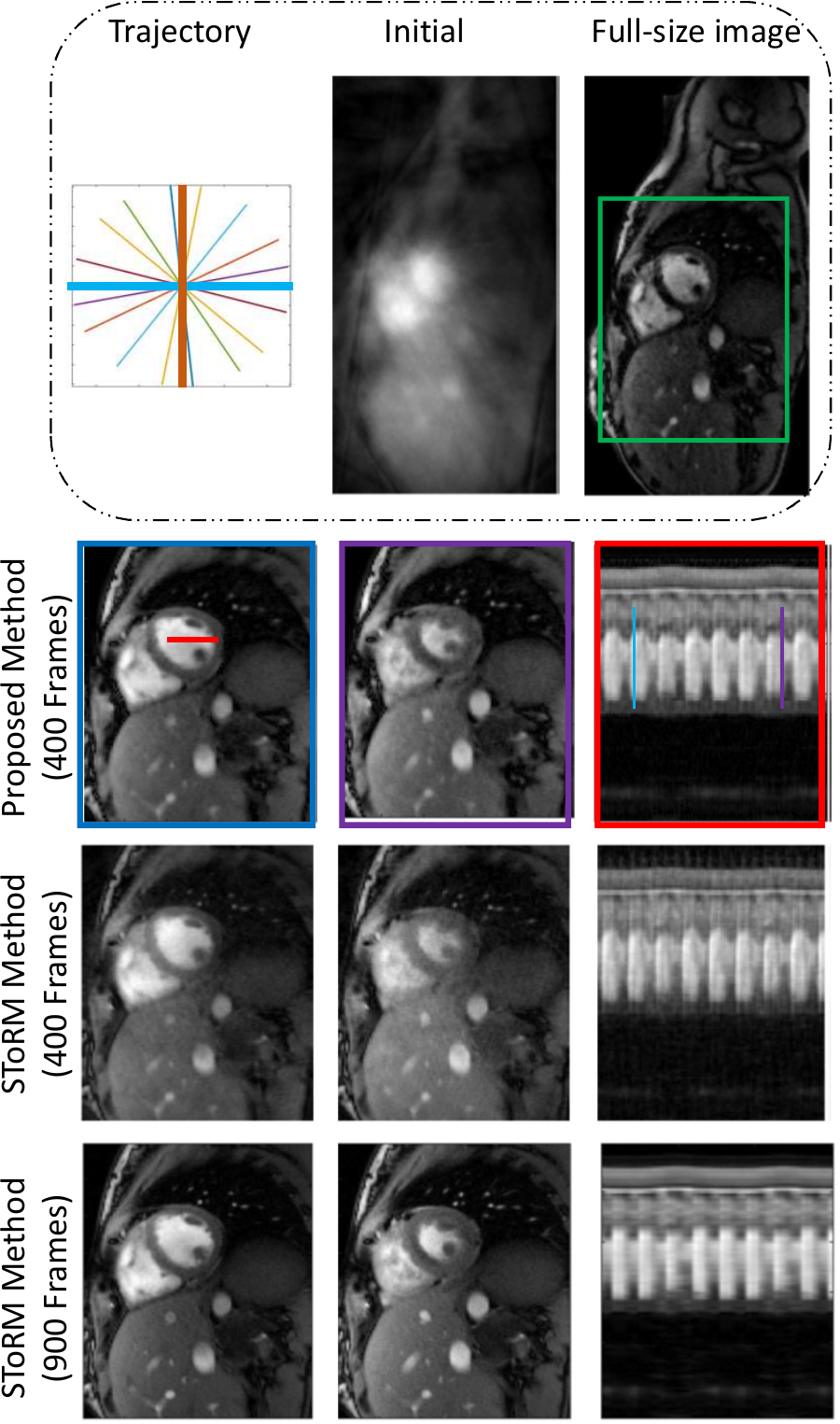}\\
	\caption {\small {First row shows radial trajectory (thick lines show navigators), initial guess and full-size image respectively. Comparisons are done on the zoomed version of full-size image around green squared box. Second row shows the result of our proposed method. Frames are picked from time points indicated by blue and purple lines. Time profile shows frames along the red color line. Results show that our method gives better spatial quality and temporal resolution with reduced aliasing artifacts.}}\vspace{-2em}
	\label{fig4}
\end{figure}
\begin{figure}[!htb]
	\centering
	\includegraphics[ scale=0.6]{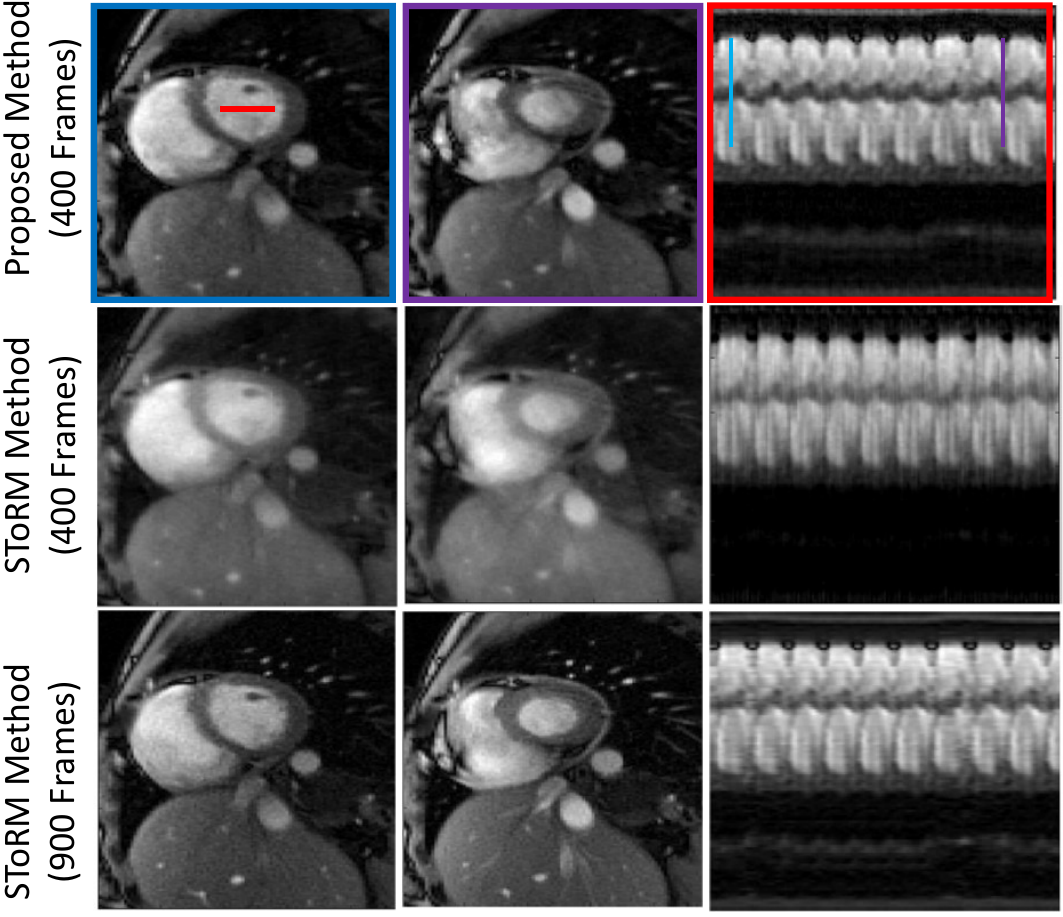}\\
	\caption {\small {Comparison of the proposed method with SToRM method on second dataset.}}
	\label{fig5}\vspace{-2em}
\end{figure}

The comparison of the proposed method against the SToRM method are shown in Fig \ref{fig4} and Fig \ref{fig5}. We observe that the SToRM scheme provides good reconstructions from 900 frames (42 seconds of data), but exhibits significant distortions when only 19 seconds of data is available. By contrast, the proposed method provides reduced aliasing artifacts and improved sharpness, when compared to the SToRM reconstruction with 400 frames.  

While the comparisons show improved reconstructions, we observe some residual blurring and alias artifacts with both 400 frame reconstructions. We will explore the additional use of spatial regularization as in \cite{MoDL-STORM} to improve the results in the future.     

\section{CONCLUSION}
\label{sec:majhead}
In this paper, we have proposed a new dynamic MRI reconstruction method based on a self-learned deep learning image prior. We have trained the denoising autoencoder using navigator data, to learn the dynamic structure in the cardiac CINE images. Then this trained network is used as an image prior to reconstruct the cardiac CINE images from highly undersampled data. Results show that this approach captures data manifold better than the kernel low-rank method. Reconstructed cardiac CINE images show the ability of our proposed method to give better spatial and temporal resolution.  
\small
\bibliographystyle{IEEEbib}
\bibliography{refs_isbi20}
\end{document}